\documentclass[referee]{aasstitre} % for a referee version
\usepackage{epsfig}
\begin{document}
  \thesaurus{09.18.1} % Stars: structure of.
   \title{Is there ERE in bright nebulae?}

   \author{Fr\'ed\'eric Zagury \inst{}  }

  % \offprints{F. Zagury}

   \institute{Department of Astrophysics, Nagoya University,
              Nagoya, 464-01 Japan 
%	      \and Institut d'Astrophysique 
%Spatiale, Universit\'e Paris Sud, Bat. 121,
%F-91405, Orsay Cedex, France
              \\
              email: zagury@a.phys.nagoya-u.ac.jp
             }

   \date{January, 2000}
   
   %Received September 15, 1996; accepted March 16, 1997}

 % \authorrunning
  \titlerunning{ERE in bright nebulae}
  
  \maketitle

 \begin{abstract}
In 1986 Witt~\& Schild studied the optical emission of selected 
bright nebulae. Comparison of the data to a model
led the authors  to conclude the nebulae have too high a level of  
$I$-band emission to be explained solely by the scattering of starlight. 
The excess is explained as a broad band $I$ emission which originates
in the nebulae. 
This phenomenum is known as Extended 
Red Emission (ERE).

Given Witt~\& Schild's data, the model they employed, and the 
estimate of observational errors, I've reached a different conclusion:  
Witt~\& Schild's observations are compatible with classical scattering and 
therefore cannot be used to prove the existence of ERE.

   \end{abstract} 
%%%%%%%%%%%%%%%%%%%%%%%%%%%%%%%%%%%%%%%%%%%%%%%%%%%%%    
 \section{Introduction}
%%%%%%%%%%%%%%%%%%%%%%%%%%%%%%%%%%%%%%%%%%%%%%%%%%%%%%%%%%%%%%%%%%%%%
In 1986, Witt and Schild (\cite{witt86}, WS hereafter) presented CCD photometric 
observations of 15 nebulae, amongst which are the brightest nebulae of the Galaxy, 
all illuminated by close B stars.

The data set assembled by WS is of great interest since it fosters 
better understanding of the properties of interstellar grains and of 
the structure of the nebulae.
The dependence of the surface brightness of a nebula with 
distance to the star, the estimate of the surface brightness per unit 
radiation field, and the color of the nebulae, narrow the search for 
better knowledge of grain properties and will indicate whether or not 
scattering alone accounts for all nebulae emission.

The surface brightness at wavelength $\lambda$, $S_\lambda$,
of a nebula assumed to scatter the light of a 
nearby star is 
proportional to the starlight flux $F^1_\lambda$ at the cloud location.
We have no direct way of knowing $F^1_\lambda$ but the ratio 
$S_{\lambda_1}/S_{\lambda_2}$ of two optical bands is 
proportional to $F^0_{\lambda_1}/F^0_{\lambda_2}$, $F^0_\lambda$ 
being the flux of the star measured on earth and corrected for reddening.
$S_\lambda$ is an observational value and $F^0_\lambda$ can be 
estimated from the observed flux of the illuminating star and the 
extinction $A_V$ in the direction of the star. 

$\Delta C^0(\lambda_1,\lambda_2)=\log(S_{\lambda_1}/S_{\lambda_2}\times 
F^0_{\lambda_2}/F^0_{\lambda_1})$ can be known with sufficient accuracy to be compared to simple models, 
such as the ones proposed by Witt (\cite{witt85}) or by Zagury, 
Boulanger and Banchet (\cite{zagury99}).
$\Delta C^0(\lambda_1,\lambda_2)$ 
depends exclusively on the optical depth of the scattering medium and on the 
properties of the grains. 
It may also depend on the structure of the 
medium if regions of different optical depths are mixed in the beam 
of the observation.

The WS paper compares the colors $\Delta C^0(B,V)$ and 
$\Delta C^0(V, I)$ of the nebulae  
 to the model proposed in Witt (\cite{witt85}).
The conclusion is summarized by figure~14 of WS' paper, which is 
reproduced here (figure~\ref{ws1}).
WS attribute the difference between the model and the observations  
to a non-scattering emission process in the $I$ band.
This Extended Red Emission (ERE) is presumed to be luminescence from a 
particular type of interstellar grain.

Before agreeing with WS' conclusion different questions 
must be clarified.

Why do most of the observational points of the WS figure~14 follow a 
curve which is parallel to the model?

How does the reddening of the stars affect $\Delta C^0$ and modify the 
relative positions of the observation values and  the model on the 
color-color plot?

According to Witt~(\cite{witt85}), no model 
can be perfect, and models of reflection nebulae usually 
meet with only limited success. 
The Witt model, which represents nebulae as 
homogenous gas masses of defined geometry, may incorrectly 
represent the nebulae.
Will small scale structure, for instance, 
modify the color forecast of an observation from the 
one predicted for a homogenous nebula?
Are the limits of the model consistent with the high column densities which likely 
prevail in the nebulae?

How do observational and data reduction errors influence WS' conclusion?
Variations of surface brightness with angular distance $\theta$ to the star, 
which roughly follows a $\theta^{-2}$ will be used to prove the abnormal behaviors 
of the I surface brightness of some nebulae. A major, 
and extremely difficult to correct, source of error is the gradient due to starlight 
diffusion in the earth's 
atmosphere which usually accompanies observations of nebulae at close 
distances to a bright star (see Zagury, Boulanger and Banchet, 
\cite{zagury99}).

Attempts to answer these questions will be found in the following 
sections.
In section~\ref{data}, WS data and argumentation are reviewed.
In section~\ref{disc}, $\Delta C^0(B,V)$ and $\Delta C^0(V, I)$ are 
extracted from the data and compared to the Witt model.
The influences of column densities higher than permitted by the low 
column density approximation of the Witt model and the structure of 
the nebulae will be discussed to reach an understanding of the 
differences between the model and the observations.

The general idea which will emerge from this study is that scattering 
can explain all the nebulae emission. 
It renders unnecessary the introduction of ERE 
carriers to explain the optical emission of this sample of nebulae.

%%%%%%%%%%%%%%%%%%%%%%%%%%%%%%%%%%%%%%%%%%%%%%%%%%%%%    
 \section{Data and WS interpretation} \label{data}
%%%%%%%%%%%%%%%%%%%%%%%%%%%%%%%%%%%%%%%%%%%%%%%%%%%%%%%%%%%%%%%%%%%%%
\begin{figure*}
\resizebox{\textwidth}{!}{\includegraphics{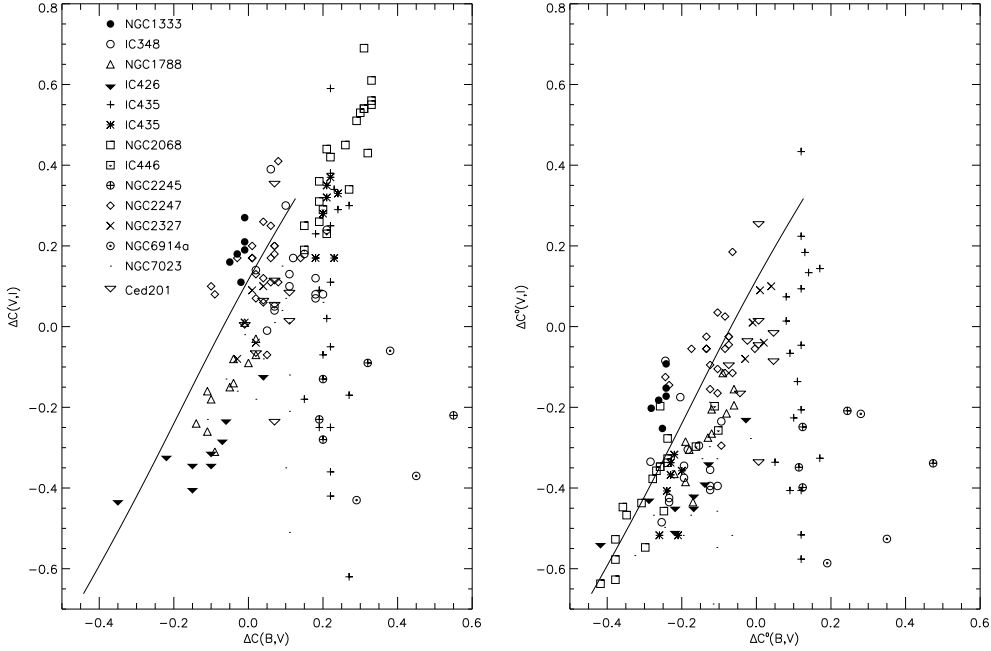}} 
\caption{\emph{left}: WS figure 15, $\Delta C(B,V)$ versus  $\Delta C(V,I)$.
The line drawn 
corresponds to $\Delta C^0(B,V)$ versus  $\Delta C^0(V,I)$ for a 
homogenous mass of gas of 
varying optical depth, computed in 
the approximation of single scattering.
\emph{right}: observed $\Delta C^0(B,V)$ versus  $\Delta C^0(V,I)$. Calculated 
from $E(B-V)$ given in WS. This plot does not figure in WS. }
\label{ws1}
\end{figure*}
%%%%%%%%%%%%%%%%%%%%%%%%%%%%%%%%%%%%%%%%%%%%%%%%%%%%%%%%%%%%%%%%%%%%%%%
For 3 to 25 chosen positions in each field, WS give 
the distance to the star, $d_\theta$, and 
$\log(S_\lambda/F_\lambda^r)$, where $S_\lambda$ and 
$F_\lambda^r$ 
are the nebula surface brightness and the star flux (not corrected 
for reddening), measured on earth, in the  $V$, $B$, $R$, $I$ bands. 
Because of the proximity of the star 
(positions are between $20$'' to $150$'' to the 
star), WS had to subtract a 
gradient due to starlight scattering in the earth's atmosphere. 

I used WS data  to reproduce their figure~15 (figure~\ref{ws1} left). 
The plot is ${\Delta}C(V,I)$ versus 
${\Delta}C(B,V)$ with 
${\Delta}C(\lambda_1,\lambda_2)=\log(S_{\lambda_1}/F_{\lambda_1}^r)-\log(S_{\lambda_2}/F_{\lambda_2}^r)$. 

The plain line in the plot is the variation with optical depth of ${\Delta}C(V,I)$ and 
${\Delta}C(B,V)$, and assumes no reddening of the illuminating star.
It was calculated from the model 
described in Zagury et al. (\cite{zagury99}) whose application is 
restricted to single scattering and is in agreement with  the Witt model. 
Both models assume the star and the 
nebula to be equally reddened by forward material -if any. 
Witt's model (see 
the equation~2 of Witt \cite{witt85}) approximates the scattered part of the 
extincted light by $g(\varphi)\omega (1-e^{-\tau})$ where $\omega$ is the albedo of 
dust grains, $g$ the phase function, $\varphi$ the angle of scattering,
and $\tau$ the optical depth of the medium.
The model applies to small optical depth mediums where single scattering dominates. 

Most 
of the observed points are below the model line, which WS interpret as 
$I$ values too large to be the result of scattering. 
They conclude that ERE accounts for $30 \% $ to $50 \% $ of the $I$ band 
surface brightness.

%%%%%%%%%%%%%%%%%%%%%%%%%%%%%%%%%%%%%%%%%%%%%%%%%%%%%    
 \section{Discussion} \label{disc}
%%%%%%%%%%%%%%%%%%%%%%%%%%%%%%%%%%%%%%%%%%%%%%%%%%%%%%%%%%%%%%%%%%%%%
%%%%%%%%%%%%%%%%%%%%%%%%%%%%%%%%%%%%%%%%%%%%%%%%%%%%%    
 \subsection{Correction for the extinction of the stars} \label{stared}
%%%%%%%%%%%%%%%%%%%%%%%%%%%%%%%%%%%%%%%%%%%%%%%%%%%%%%%%%%%%%%%%%%%%%
I will introduce:
%%%%%%%%%%%%%%%%%%%%%%%
\begin{equation}
{\Delta}C^0(\lambda_1,\lambda_2)=
\log(S_{\lambda_1}/F_{\lambda_1})-\log(S_{\lambda_2}/F_{\lambda_2})    
    \label{eq:ws1}
\end{equation}¥
%%%%%%%%%%%%%%%%%%%%%%%%%%%%%%
$F_\lambda$ is the star flux at wavelength $\lambda$, measured on earth and corrected for 
reddening. 

$\Delta C^0$ so defined is a possible definition for the color of a 
nebula. 
A significant difference exists between the traditional meaning of 
color and $\Delta C^0$: when increasing the optical depth of a 
low density medium,  $B-V$, $B-I$ and $V-I$ colors increase 
while ${\Delta}C^0(B,V)$, ${\Delta}C^0(B,I)$, ${\Delta}C^0(V,I)$ decrease. 

${\Delta}C^0(V,I)$ and ${\Delta}C^0(B,V)$ anticipated values for a medium 
with little reddening are ${\Delta}C^0(V,I)_{max}=0.34$ and 
${\Delta}C^0(B,V)_{max}=0.13$ ($A_B/A_V=1.34$ 
and $ A_I/ A_V=0.45$, Cardelli et al. \cite{cardelli89}). Extinction 
effects in mediums with 
increasing optical depths can only decrease (redden) ${\Delta}C^0(V,I)$ 
and ${\Delta}C^0(B,V)$, so that one should always have: 
${\Delta}C^0(V,I)<{\Delta}C^0(V,I)_{max}$, 
and ${\Delta}C^0(B,V)<{\Delta}C^0(B,V)_{max}$, if scattering alone is 
considered. 

The model curve of figure~\ref{ws1} is the theoretical color-color 
plot (${\Delta}C^0(V,I)$,  ${\Delta}C^0(B,V)$) for a medium of low 
optical depth.
The model assumes that starlight is not reddened between the star 
and the nebula.
Starlight extinction between the star and the scattering volume will 
redden the visible emission of the nebula and move the model curve 
down and to the left, thus approaching the observations.

${\Delta}C^0(V,I)$ and  ${\Delta}C^0(B,V)$ can be estimated from
${\Delta}C(V,I)$ and  ${\Delta}C(B,I)$ in the following manner.
At optical wavelengths, 
$A_\lambda$ is a linear function of $1/\lambda$ (Cardelli, Clayton~\& 
Mathis  \cite{cardelli89}, Rieke~\& Lebofsky~\cite{rieke85}). 
Hence:
%%%%%%%%%%%%%%%%%%%%%%%
\begin{eqnarray}
   A_{\lambda_1}-A_{\lambda_2}&\,=\,&\frac{E(B-V)}{\frac{1}{\lambda_B}-\frac{1}{\lambda_V}}
   \left(\frac{1}{\lambda_1}-\frac{1}{\lambda_2}\right) \nonumber \\
   &\,=\,& 2.2E(B-V)\left(\frac{1\mu\rm m}{\lambda_1}-\frac{1\mu\rm m}{\lambda_2}\right) \label{eq:al}
\end{eqnarray}¥
%%%%%%%%%%%%%%%%%%%%%%%%%%%%%%%%
Use of relation~\ref{eq:al} in the definition of ${\Delta}C^0$ gives:
%%%%%%%%%%%%%%%%%%%%%%%
\begin{eqnarray}
{\Delta}C^0(\lambda_1,\lambda_2)&\,=\,& {\Delta}C(\lambda_1,\lambda_2)- 
0.9E(B-V)\left(\frac{1\mu\rm m}{\lambda_1}-\frac{1\mu\rm m}{\lambda_2}\right)
    \label{eq:ws2} \\
{\Delta}C^0(B,V)&\,=\,& {\Delta}C(B,V)-0.4E(B-V)
    \label{eq:bv} \\
{\Delta}C^0(V,I)&\,=\,& {\Delta}C(V,I)-0.6E(B-V)
    \label{eq:vi}     
\end{eqnarray}¥
%%%%%%%%%%%%%%%%%%%%%%%%%%%%%%
Figure~\ref{ws1}, left, plots the observed values of 
${\Delta}C(V,I)$ and  ${\Delta}C(B,I)$ for the nebulae WS have 
considered.
Most points 
follow a curve parallel to the model curve but shifted down from it.

A few points, mainly observations of NGC2068, 
have ${\Delta}C(V,I)$ greater than ${\Delta}C^0(V,I)_{max}$.
Roughly half of ${\Delta}C(B,V)$ values of the plot are significantly above $0.13\,$mag. 
This is to be attributed to the  reddening of the stars which decreases  
${\Delta}C^0(B,V)$ and ${\Delta}C^0(V,I)$ according to 
equations~\ref{eq:bv} and \ref{eq:vi}. 
Correction for the reddening of the stars will shift all the 
observations down by $0.6E(B-V)\,$mag and
 to the left by $0.4E(B-V)\,$mag.
$E(B-V)$ is given in table~1 of WS.

The color-color diagram (${\Delta}C^0(V,I)$, ${\Delta}C^0(V,I)$) is represented on 
the right hand plot of figure~\ref{ws1}.
Most nebulae observations are now scattered close to the model curve. 

Different reasons may explain departures from the model line and the 
tendency for the nebulae to remain under the curve.
Error bars explain part of the difference between the model and the 
observations.
The Witt model is restricted to low column 
densities since it requires single scattering and low optical depth. 
What is the effect of an increase of column density?
The nebulae of the Witt model must be homogenous enough 
for the medium sampled by the beam of the observation to be described by one 
value of the optical depth ($\tau_0$ in Witt's model).
How will small scale structure modify WS' conclusions?
%%%%%%%%%%%%%%%%%%%%%%%%%%%
\subsection{The limits of the Witt model} \label{modlim}
%%%%%%%%%%%%%%%%%%%%%¥
The limit of validity of the model curve (figure~\ref{ws1}) is that single 
scattering dominates for each of the optical bands.
It is satisfied if the scattering volumes are proven to be 
homogenous (can be described by a single optical depth) and have a low $A_V$ value.
According to Witt (\cite{witt85}) the model is valid for 
$A_V\sim \tau<\tau_{lim}=0.6$.
Since this condition must be satisfied for all the optical bands, it 
implies: $A_V\sim \tau_{V}<0.6$, hence $A_I\sim 0.5A_V < 0.3$.

WS' representation of the medium surrounding the stars (\S~III~b 
in WS),  by a homogenous 
spherical nebula of radial optical depth $ A_V\sim0.5\,$mag, is within 
the limits of the single scattering model but is questionable.
The selection criteria of nebulae with high surface brightnesses in largely extincted 
regions, some of which are starforming 
regions, supports the idea of high column densities and perhaps a wide 
range of structures with different optical depths along each line of sight. 

High column densities in the nebulae are also indicated by the extinction 
in the direction of the stars. $A_V$ in the stars direction should be the radial 
optical depth to use in the Witt model.
Nearly all the stars in the sample have large reddening.
$E(B-V)$ for all stars except one, the illuminating star of IC426,
is greater than $0.2$, with a mean value $\sim 0.7$ (WS, 
table~1). 
With a minimum $R_V=A_V/E(B-V)$ of $3$ (Cardelli et 
al.~\cite{cardelli89}), the $A_V=1.08\tau_\star$ value found for the nebulae will range 
from $0.6$ to $4.2$, with an average value of $2$, far above 
$\tau_{lim}$ and above the radial optical depths adopted in WS. 

Observations do not agree with WS' concept of the 
nebulae.
Column densities must be higher than $A_V=0.5$.
The effect on the color-color plot of such high column densities, out of the range of 
validity of the single scattering model, is investigated in 
section~\ref{coldens}.

A clumpy medium made of cells of different column densities, and much 
smaller than the resolution of the observations, may be an 
alternative representation of the homogenous mass of gas adopted in WS.
It will modify (section~\ref{sss}) the color of the nebulae and can 
explain the high surface brightnesses observed in all the optical bands.
%%%%%%%%%%%%%%%%%%%%%%%%%%
\subsection{Effect of an increase of the column density on the model 
curve} \label{coldens}
%%%%%%%%%%%%%%%%%%%%%¥
If the column density of the medium which is observed is increased, 
the single scattering approximation will stop being valid in the $B$ and 
in the $V$ bands before the $I$ band.
At low column density ($A_V \ll 1$), scattering in the $I$ band will be 
 $\sim 3$ times less (Cardelli et al. \cite{cardelli89})
than in the $B$ band.
Increasing the column density will see the domination of absorption effects 
in the $B$ band while scattering in the $I$ band will become important. 
For $A_V \sim 3$, we have $A_I\sim 1.35$ and $A_B \sim 4$.
Given these values, scattering will be minimal in the $V$ and $B$ 
bands, most of the $B$ and $V$ light will be absorbed.
It is close to its maximum in the $I$ band.
$B$ and $V$ surface brightnesses will be close to $0$ and $I$ surface 
brightness close to its maximum.

Hence, the net effect of an increase of column density will be to shift the 
model points of figure~\ref{ws1} down, as observed. 
The points are also shifted to the left, but the effect is less important
(section~\ref{stared}, equations~\ref{eq:bv} and \ref{eq:vi}).  

The model curve of figure~\ref{ws1} is a limit 
which separates two regions.
All observations should lie in the region under the curve.
The few points above the model curve will be explained by the error 
margin of the observations.

%%%%%%%%%%%%%%%%%%%%%%%%%%%
\subsection{Effect of the small scale structure} \label{sss}
%%%%%%%%%%%%%%%%%%%%%¥
The ambiguity between the high column densities which exist in 
the nebulae and the observed high surface brightnesses in all bands 
can be removed if the emission in the various bands is not exactly produced by the 
same interstellar regions on the same line of sight. High $I$ band surface 
brightness arises from regions with relatively high optical depth 
($A_V>1$) which will absorb $V$ and $B$ starlight and produce little 
emission in those bands. The $B$ and $V$ band emissions should come from 
regions of the same line of sight, with lower $A_V$, as in the WS model. 
Those regions will 
give little scattering in the $I$ band. If this is the case, one cannot 
expect the observations to match the uniform mass of gas model.

The contribution of structures with high $A_V$ to the
emission in the $I$-band will contribute to the drop in  ${\Delta}C^0(V,I)$ observed for 
some points in the right plot of figure~\ref{ws1}. 

The probability that the observed nebulae are structured at small scales 
is strengthened by the beam size of WS observations. WS images' 
resolution is not better than $10$'' and the average 
distance to the nebulae is, according to WS' table~I, $600\pm 300\,$pc. 
Thus, the spatial  resolution of the observations is at least 
$0.03\,$pc, a large value ($3$ orders of magnitude) compared to the 
possible interstellar clouds scale ($\sim$ a few tens of A.U., 
Falgarone et al.~\cite{falgarone}, Zagury et al.~\cite{zagury99}). 

%%%%%%%%%%%%%%%%%%%%%%%%%%%
\subsection{Error bars}\label{err}
%%%%%%%%%%%%%%%%%%%%%¥
%%%%%%%%%%%%%%%%%%%%%%%%%%%%%%%%%%%%%%%%%%%%%%%%%%%%%%%%%%%%%%%%%%%%%
\begin{figure*}
\resizebox{\textwidth}{!}{\includegraphics{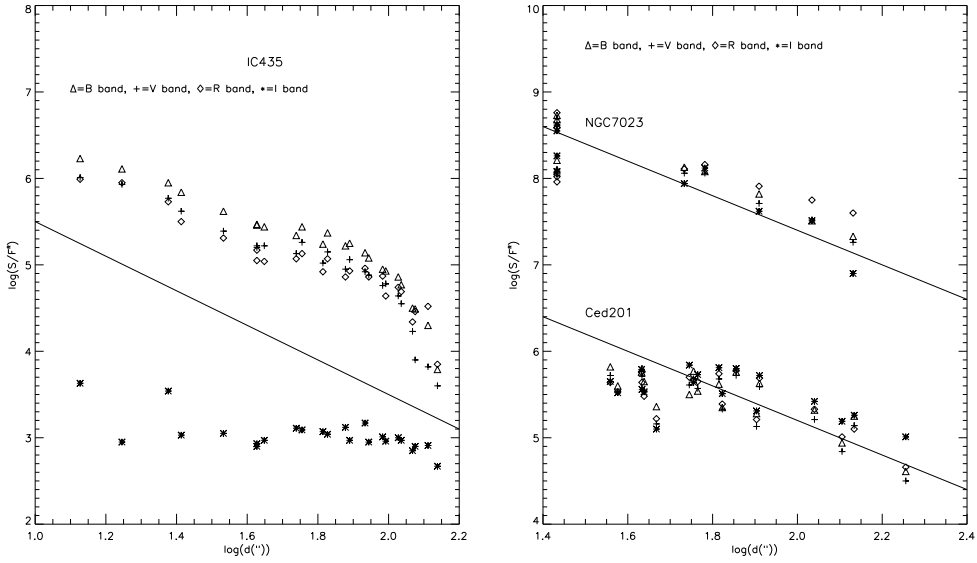}} 
\caption{Plot of $\log(S_{\lambda}/F_{\lambda})$ versus 
distance to the illuminating star for the $B$, $V$, $R$, $I$ bands and for 
$3$ nebulae. The $I$ 
band values of IC435 (left plot, asterix) are nearly constant and do not follow the 
variations of the other band, as in most other fields. A plausible 
interpretation is the gradient subtracted from $S_{\lambda}$,  which may 
introduce a variable offset. } 
\label{wsdis}
\end{figure*}
%%%%%%%%%%%%%%%%%%%%%%%%%%%%%%%%%%%%%%%%%%%%%%%%%%%%%%%%%%%%%%%%%%%%%%%
Most observations (figure~\ref{ws1}, right) lie under the model line 
and at close distance to it, which can be justified either by 
higher column densities than permitted by the model or by the small scale 
structure of the medium. Error bars will explain the extreme position 
on the plot of some observations.

The uncertainty of $E(B-V)$ introduces an error on ${\Delta}C^0(B,V)$ and 
on ${\Delta}C^0(V,I)$. 
The error on $E(B-V)$ is probably less than $0.1$ and may produce a 
similar error in  ${\Delta}C^0(B,V)$ and ${\Delta}C^0(V,I)$. 

Error bars on the observed values, ${\Delta}C(B,V)$ and ${\Delta}C(V,I)$,
may be larger than WS estimates due to the 
atmospheric gradient substracted from each observation. The same 
problem was dealt with in Zagury et 
al. (\cite{zagury99}): a precise subtraction of the atmospheric 
gradient was found to be impossible to arrive at. 

%%%%%%%%%%%%%%%%%%%%%%%%%%%
\subsubsection{NGC1333 and NGC2247}\label{ngc1333}
%%%%%%%%%%%%%%%%%%%%%¥
The singular positions above the model line of NGC1333  
and of some of NGC2247 observations must be due to an error in the 
data. 
For NGC2247, the error, given in WS, $\rm \sim 
\pm 0.04\,to\,\pm 0.12$ for ${\Delta}C(B,V)$ and ${\Delta}C(V,I)$, 
 is enough to bring all points back under the model line. 

Errors given for NGC1333 are lower, of order $\pm 0.04$, and may have 
been slightly underestimated. 
Along with a possible 
overestimation of the $E(B-V)$ value of BD$+30^{\circ}549$, the 
illuminating star of NGC1333, error bars also explain
the extreme positions of NGC1333 observations in the right plot of 
figure~\ref{ws1}.

%%%%%%%%%%%%%%%%%%%%%%%%%%%
\subsubsection{IC435, NGC2245 and NGC6914a}\label{ic435}
%%%%%%%%%%%%%%%%%%%%%¥
Three nebulae, IC435, NGC2245, NGC6914a are far under the model curve. 
Some of these points have ${\Delta}C^0(B,V)>{\Delta}C^0(B,V)_{max}$ which must 
indicate error bars larger than WS' estimates.

Figure~\ref{wsdis} displays  $\log(S_{\lambda}/F_{\lambda})$ versus $\log 
d_\theta$ for some stars. 
For most nebulae the surface brightness decreases when moving 
away from the star, in all the optical bands. 
The decrease follows a $1/d_\theta ^2$ law for the
fields NGC1788, IC446, NGC2247, NGC2327, NGC7023, and Ced201 
(figure~\ref{wsdis}, right plot). 
Similar variations of $\log(S_{\lambda}/F_{\lambda})$ versus $\log d_\theta$ 
are observed in most fields and all optical bands. 
Important differences also arise, especially in the $I$ 
band. 

A particularly neat example is IC435 where $B$, $V$, $R$ surface 
brightnesses decrease as $1/d_\theta^2$, while $I$ surface brightness 
remains constant (figure~\ref{wsdis}, left plot). 
Log$(S_{\lambda}/F_{\lambda}^r)$ values for IC345 are nearly equal in the $B$, $V$, $R$ 
bands (figure~\ref{ws1}) at all positions. 
The peculiar behaviour of IC435 in the $I$-band (figure~\ref{wsdis}, 
left) is most probably explained by a large error in $I$ surface brightness.
The error on 
${\Delta}C(B,V)$ and ${\Delta}C(V,I)$ for IC435, estimated by WS, is of order $\pm 0.1$. 
An error of $\pm 0.2$ 
to $\pm 0.3$ seems to be more likely for ${\Delta}C(V,I)$.

Like IC435, the $I$-band variations (not represented here) of NGC2245 and NGC6914a do not 
follow the $B$, $V$, and $R$ band variations.  Errors for these two stars, 
given in WS, are substantially higher than for IC435. They reach $0.2$, 
a value comparable to the one infered for IC435.
%%%%%%%%%%%%%%%%%%%%%%%%%%%%%%%%%%%%%%%%%%%%%%%%%%%%%    
 \section{Conclusion}
%%%%%%%%%%%%%%%%%%%%%%%%%%%%%%%%%%%%%%%%%%%%%%%%%%%%%%%%%%%%%%%%%%%%%
In this paper I tried to show that WS analysis of the emission of 
bright nebulae is not conclusive.
The observations are compatible 
with starlight scattering as the only process involved.

The first part of WS' paper is dedicated to the comparison of the 
color of the nebulae with Witt's model, which applies to low column 
density nebulae where single scattering in all optical bands dominates.
The curve which represents this model in a (${\Delta}C^0(B,V)$, 
${\Delta}C^0(V,I)$) plot is to be considered 
as a limit under which all observations should be. 

Correction for the reddening of the stars will scatter most 
observations close to the model line, where they have a tendency to remain under 
the line, as expected.

Three reasons explain the remaining differences between the model and the 
observations. 

A deeper analysis of the error margin of the 
observations explains the largest differences between the Witt model and WS' 
observations. 
The main source of error is the gradient of  starlight scattered in the 
earth's atmosphere which needs to be subtracted from the observations.

The general drop of the observations under the model curve
can be attributed to column densities higher than tolerated by the 
single scattering model.
Fitting the observations to the model in the $B$ or the $V$ band, 
as is done in WS, will give low $A_I$ and low $I$ surface brightness values for the nebulae, 
incongruous with the large reddening of the stars. 
The $A_V$ values in the direction of 
the stars are between $1$ and $4$, with a mean value of $2$,
above the maximum optical depth ($\sim 0.6$) supported by the model.

High surface brightness in the I-band does not prove ERE, but indicates 
optical depths in the $B$ and $V$ bands higher than 
can be supported by the model.
The expected position of these observations on the color-color plot will 
be under the curve, as it is observed.

The small scale structure of the 
nebulae will also heighten differences between the Witt 
model and the observations.  
A structured medium permits high surface brightnesses in all bands 
since the bulk of the emission at two different wavelengths need not arise 
from the same interstellar 
structures.
Small scale structure in the sample of nebulae presented in WS is 
probable because of the large beam of the observations and the high 
column densities in the nebulae.

The Witt model, which represents the nebulae 
as spheres of constant density centered on the star, never fits the 
$V$ band observations variation with distance to the central star.
But up to what point is it reasonable to expect the observations to 
match the model? 
When the observations do not match the representation of 
a nebula proposed in WS
shouldn't the model be questioned first, before concluding that
interstellar grains have special properties which vary from cloud to cloud?

The many different aspects discussed in this paper indicate that WS' observations
can be explained by scattering 
effects only. While it certainly is possible to interpret the data with 
excess of red emission, this interpretation requires a 
constrained and 
limited model which is probably a poor representation of reality and 
whose reliability the authors have yet to prove.

\end{document}